\pdfoutput=1

\documentclass[12pt]{article}

\usepackage{graphicx}
\graphicspath{{fig/}}

\usepackage[skip=3pt]{subcaption}  

\usepackage{amsmath,amssymb}
\usepackage{hyperref}

\newcommand\pubnumber{}
\newcommand\pubdate{November 6, 2014}

\def\Title#1{\begin{center} {\Large #1 } \end{center}}
\def\Author#1{\begin{center}{ \sc #1} \end{center}}
\def\Address#1{\begin{center}{ \it #1} \end{center}}

\newcommand\pubblock{\rightline{\begin{tabular}{l} \pubnumber\\
         \pubdate  \end{tabular}}}
\newenvironment{Abstract}{\begin{quotation}  }{\end{quotation}}
\newenvironment{Presented}{\begin{quotation} \begin{center} 
             PRESENTED AT\end{center}\bigskip 
      \begin{center}\begin{large}}{\end{large}\end{center} \end{quotation}}

\hypersetup{pdfauthor={Andrey Popov on behalf of the CMS collaboration}, %
 pdftitle={Measurements of single top-quark production in pp collisions by the CMS experiment}}

\begin{document}
\begin{titlepage}
\pubblock

\vfill
\Title{Measurements of single top-quark production in $pp$~collisions\\ by the CMS experiment}
\vfill
\Author{Andrey Popov on behalf of the CMS collaboration}
\Address{Universit\'{e} catholique de Louvain, Louvain-la-Neuve, Belgium\\
Also at Lomonosov Moscow state university, Moscow, Russia}
\vfill
\begin{Abstract}
A summary of studies of electroweak top-quark production performed by the CMS collaboration is presented. The results include measurements of production cross sections, extraction of the value of $|V_{tb}|$, determination of $W$-boson helicity in top-quark decays and top-quark spin asymmetry, and a search for anomalous couplings in the $Wtb$ vertex. No deviations from predictions of the standard model are found.
\end{Abstract}
\vfill
\begin{Presented}
The 8$^\text{th}$ international workshop on the CKM unitarity triangle (CKM 2014)\\ Vienna, Austria, September 8--12, 2014
\end{Presented}
\vfill
\end{titlepage}
\def\thefootnote{\fnsymbol{footnote}}
\setcounter{footnote}{0}

\section{Introduction}

\begin{figure}[htb]
\centering
 \subcaptionbox{}{\includegraphics[width=0.2\textwidth]{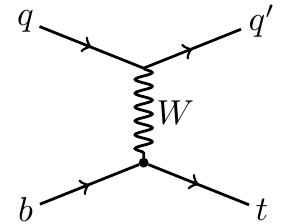}}\quad
 \subcaptionbox{}{\includegraphics[width=0.4\textwidth]{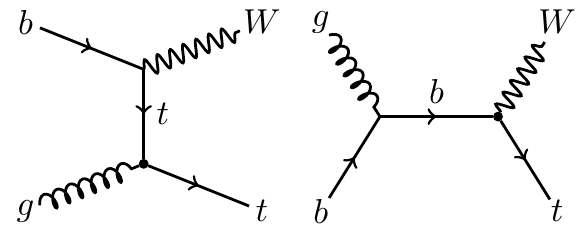}}\quad
 \subcaptionbox{}{\includegraphics[width=0.2\textwidth]{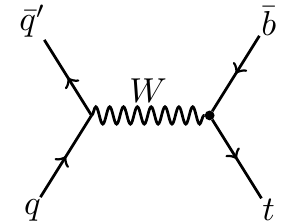}}
\caption{Representative Feynman diagrams for electroweak top-quark production: $t$-channel~(a), $tW$-channel~(b), and $s$-channel~(c).}
\label{FigDiagrams}
\end{figure}

Top~quark is the most massive known elementary particle and as such is often speculated to have a special sensitivity to possible physics beyond the standard model (SM). At LHC it is produced predominantly in pairs via the strong interaction, but electroweak production of single top~quarks is not negligible. At leading order it can be classified into three production channels, depending on the virtuality of the involved $W$~boson, as shown in Fig.~\ref{FigDiagrams}.

This paper summarises experimental studies of single top-quark production performed with the CMS detector~\cite{Chatrchyan:2008aa}. It does not cover however flavour-changing neutral currents discussed in~\cite{LiuCKM14} or searches of new particles in final states involving single top quarks.

\section{Cross sections and \texorpdfstring{\boldmath$|V_{tb}|$\unboldmath}{|Vtb|}}

Measurements in all three production channels were performed. Although applied event selections depend on the target channel, the presence of a $b$-tagged jet, which is a jet that is identified as stemming from hadronisation of a $b$~quark, is required and only semileptonic decays of the top~quark with muons or electrons in the final state are considered.

\subsection{Measurement in the \texorpdfstring{\boldmath$t$}{t}~channel}

At LHC production in the $t$~channel is the dominant of the three modes, and as such was studied first. A prominent feature of this process is the presence of a light-flavour recoil jet (denoted $q'$ in Fig.~\ref{FigDiagrams}a) that is often forward. Thus, the experimental signature of signal events is exactly one isolated muon or electron, a moderate missing $E_\text{T}$ due to a neutrino, and two jets, only one of which is $b$-tagged. Main backgrounds are top pair and $W$~boson production.

The production cross section was measured at centre-of-mass energies of 7 and 8\,TeV. At $\sqrt s = 7$\,TeV three independent analyses were performed~\cite{Chatrchyan:2012ep}. 
The first approach exploited the characteristic non-central distribution in pseudorapidity of the recoil jet~$\eta_{j'}$ and additionally enhanced purity of the event selection requiring that the reconstructed top-quark mass is close to the nominal value. It was complemented by two more complex approaches based on methods of multivariate analysis (MVA). Results of the three analyses were compatible with each other and were combined yielding a cross section of $67.2 \pm 6.1$\,pb.

\begin{figure}[htb]
\centering
 \includegraphics[width=0.49\textwidth]{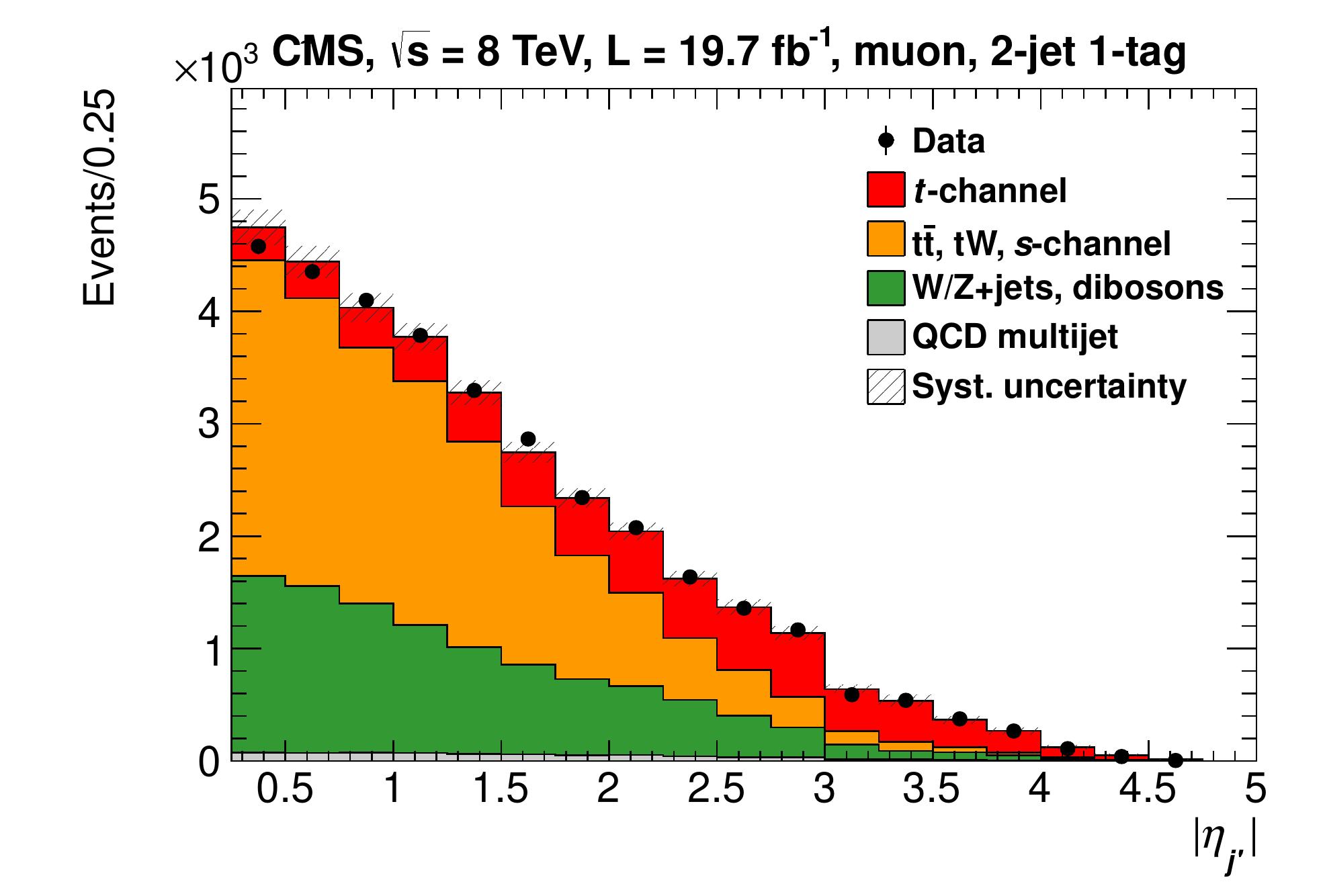}
 \includegraphics[width=0.49\textwidth]{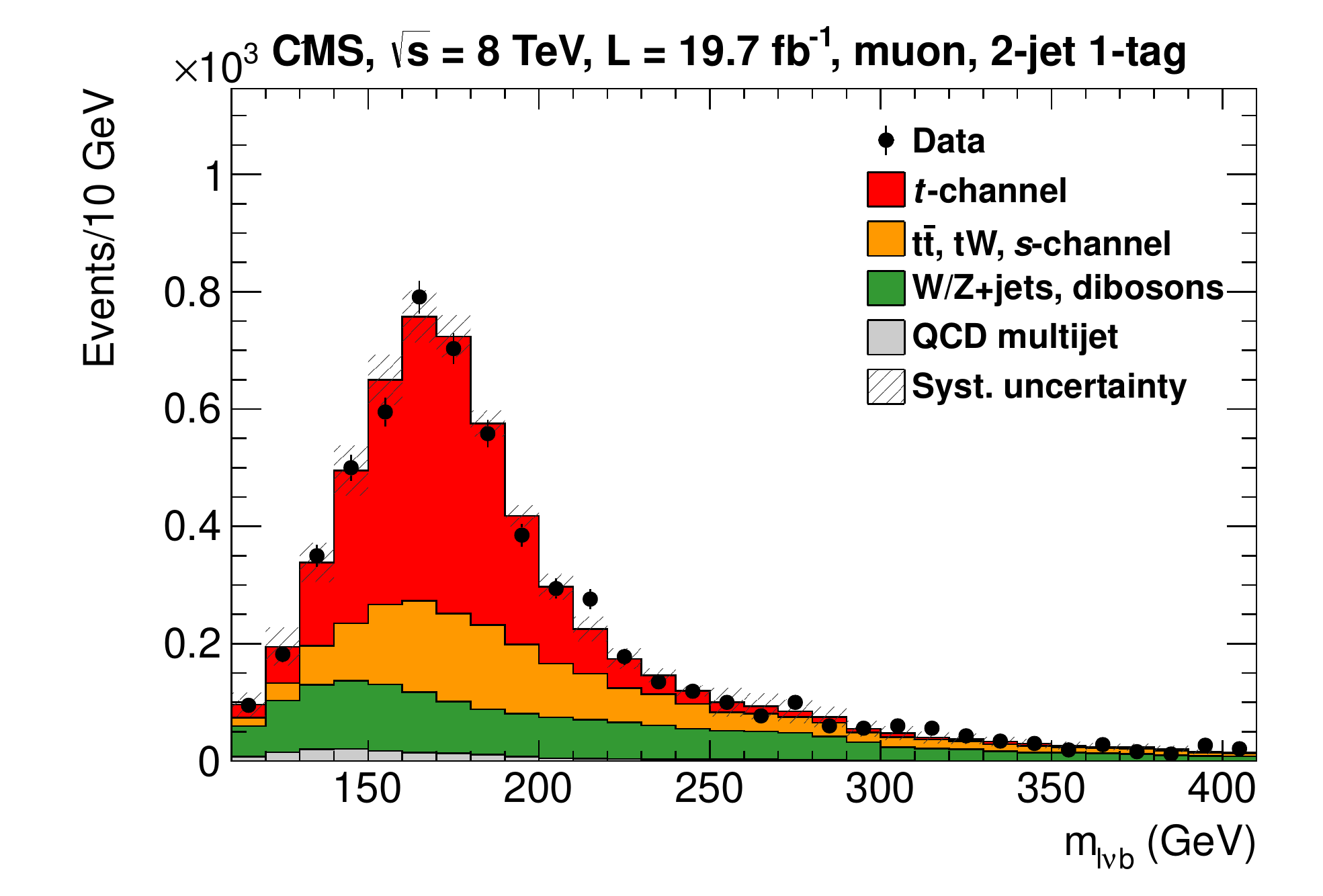}
\caption{Post-fit distributions of the pseudorapidity of the recoil jet (left) and the reconstructed top-quark mass in the region $|\eta_{j'}| > 2.5$ (right) in the 8\,TeV $t$-channel measurement, muon channel.}
\label{FigTChanXSec}
\end{figure}

At $\sqrt s = 8$\,TeV the analysis based on pseudorapidity of the recoil jet was re-optimised and repeated~\cite{Khachatryan:2014iya}. 
Resulting post-fit distributions of $\eta_{j'}$ and reconstructed top-quark mass are shown in Fig.~\ref{FigTChanXSec}. The production cross section was measured to be $83.6 \pm 7.7$\,pb. Separate measurements for top quarks and anti-quarks were also performed, giving a ratio $\sigma_t / \sigma_{\bar t} = 1.95 \pm 0.21$. In addition, a ratio to the 7\,TeV cross section was determined to be $\sigma_\text{8\,TeV} / \sigma_\text{7\,TeV} = 1.24 \pm 0.14$.

\subsection{Observation of the \texorpdfstring{\boldmath$tW$}{tW}~channel}

Measurements in the $tW$~channel were performed with a dilepton selection, requiring two leptons of opposite electric charge: $\mu^+\mu^-$, $e^+e^-$, or $\mu^\pm e^\mp$. In addition, selected events had to contain exactly one $b$-tagged jet. The dominant background is $t\bar t$~production.

Measurements were conducted at $\sqrt s = 7$ and 8\,TeV~\cite{Chatrchyan:2012zca, Chatrchyan:2014tua}. 
In both cases MVA-based discriminators were exploited, with simpler approaches adopted as cross-checks. At 7\,TeV a signal significance of $4.0\,\sigma$ was observed, with an expectation of $3.6^{+0.8}_{-0.9}\,\sigma$; the cross section was measured to be $16^{+5}_{-4}$\,pb. With the 8\,TeV data the first observation of this channel was reported, with an observed (expected) signal significance of $6.1\,\sigma$ ($5.4 \pm 1.4\,\sigma$). A cross section of $23.4 \pm 5.4$\,pb was measured.

\subsection{Search of the \texorpdfstring{\boldmath$s$}{s}~channel}

The $s$-channel production has the smallest cross section at LHC. A search for it was performed at $\sqrt s = 8$\,TeV~\cite{CMS:2013fmk}. 
The event selection required the presence of two $b$-tagged jets in addition to an isolated muon or electron. Signal events were identified with the help of an MVA, and a special procedure was deployed to constrain the dominant $t\bar t$~background in situ.
A signal significance of $0.7\,\sigma$ was observed, with an expectation of $0.9^{+1.0}_{-0.9}\,\sigma$, and an upper limit of 11.5\,pb at 95\% confidence level was set.

\subsection{Extraction of \texorpdfstring{\boldmath$|V_{tb}|$}{|Vtb|}}

The absolute value of the $V_{tb}$ element of the CKM matrix can be extracted from a measurement of the single-top production cross section. Under assumptions that $|V_{td}|, |V_{ts}| \ll |V_{tb}|$ and the $Wtb$~vertex is purely left-handed, $|V_{tb}| = \sqrt{\sigma_\text{exp} / \sigma_{|V_{tb}| = 1}}$, where $\sigma_\text{exp}$ is the measured cross section and $\sigma_{|V_{tb}| = 1}$ is the theoretical expectation with $|V_{tb}| = 1$. The best precision on $|V_{tb}|$ was achieved by the combination of the $t$-channel results at 7 and 8\,TeV~\cite{Khachatryan:2014iya}. A value of $|V_{tb}| = 0.998 \pm 0.038 \text{\,(exp.)} \pm 0.016 \text{\,(theo.)}$ was determined, which can be translated into a lower limit $|V_{tb}| > 0.92$ at the 95\% confidence level assuming that $0 \leqslant |V_{tb}| \leqslant 1$.

\section{Properties}

Relatively high cross section and purity of event selection in the $t$~channel allow studying single-top events in detail, and several measurements of properties were performed in this channel.

\subsection{\texorpdfstring{\boldmath$W$}{W}-boson helicity in single-top topology}

Polarisation of the $W$~boson in the $t\to bW$ decay encodes information about the chiral structure of the $Wtb$ vertex and can be probed by studying the distribution in the angle~$\theta$ between the momentum of the charged lepton in $W\to l\nu$ and the opposite of the direction of the $b$~quark, calculated in the rest frame of the $W$~boson. From a fit to $\cos\theta$ one can extract fractions of $W$~bosons produced with left-handed ($F_L$), right-handed ($F_R$), or longitudinal ($F_0$) helicity.

\begin{figure}[htb]
\centering
 \includegraphics[width=0.4\textwidth]{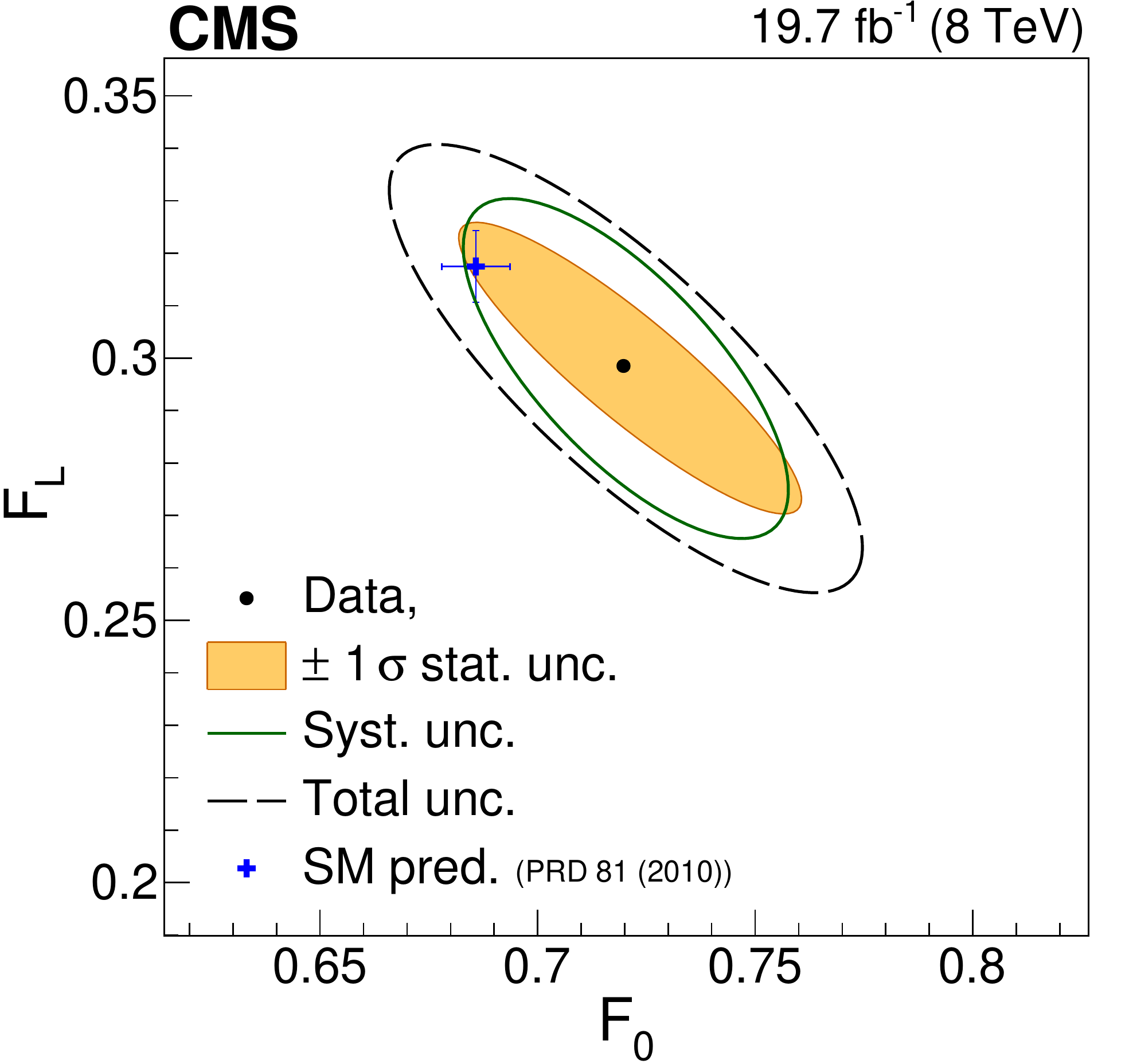}\quad
 \includegraphics[width=0.38\textwidth]{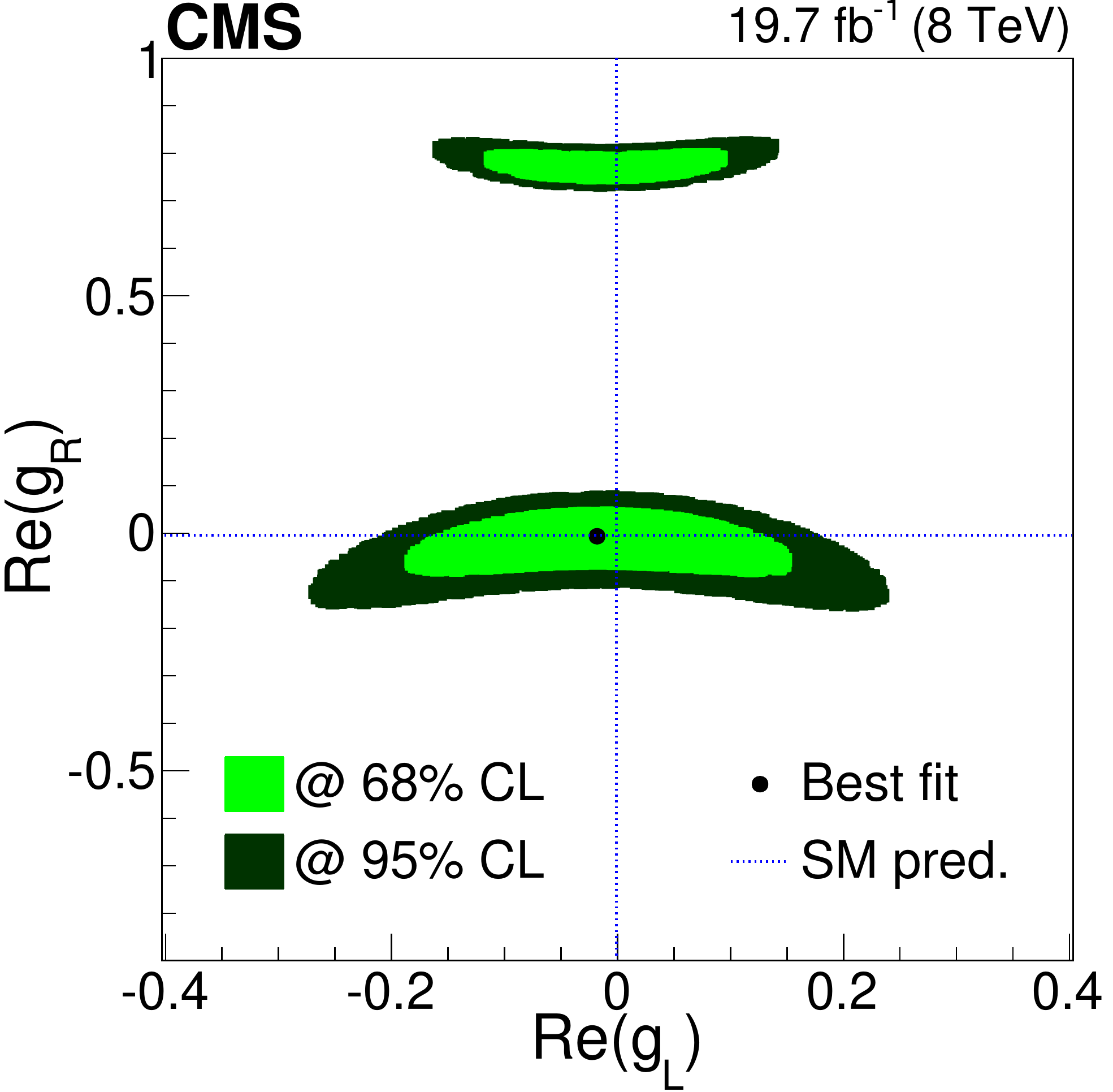}
\caption{Measured $W$-boson helicity fractions~(left) and their translation into exclusion limits on anomalous tensor couplings in the $Wtb$ vertex~(right).}
\label{FigWHelicityResults}
\end{figure}

The $W$-helicity fractions were measured at $\sqrt s = 8$\,TeV~\cite{Khachatryan:2014vma} utilising both single-top and $t\bar t$ events that pass the selection. The results are shown in Fig.~\ref{FigWHelicityResults}.

\subsection{Top-quark polarisation}

In SM, top~quarks are largely polarised along the direction of the recoil quark. This polarisation is reflected in the distribution over the angle $\theta^*$ between the charged lepton in $t\to bl\nu$ and the recoil quark, calculated in the rest frame of the top~quark. A top-quark spin asymmetry defined as
\begin{equation}
 A_l = \frac{N(\theta^* < \pi/2) - N(\theta^* > \pi/2)}{N(\theta^* < \pi/2) + N(\theta^* > \pi/2)},
 \label{EqAsymmetry}
\end{equation}
where $N$ denotes the number of single-top events in which $\theta^*$ is larger or smaller than $\pi/2$, was measured~\cite{CMS:2013rfa} with 8\,TeV data. A sample enriched in $t$-channel single-top events was selected with the help of an MVA. Remaining contribution from backgrounds was subtracted from the distribution of selected events in $\cos\theta^*$, and the distribution was unfolded to the parton level. Then the asymmetry~\eqref{EqAsymmetry} was calculated from it, yielding a value of $A_l = 0.41 \pm 0.17$.

\subsection{Anomalous couplings in the \texorpdfstring{\boldmath$Wtb$}{Wtb}~vertex}

\begin{figure}[htb]
\centering
 \includegraphics[width=0.45\textwidth]{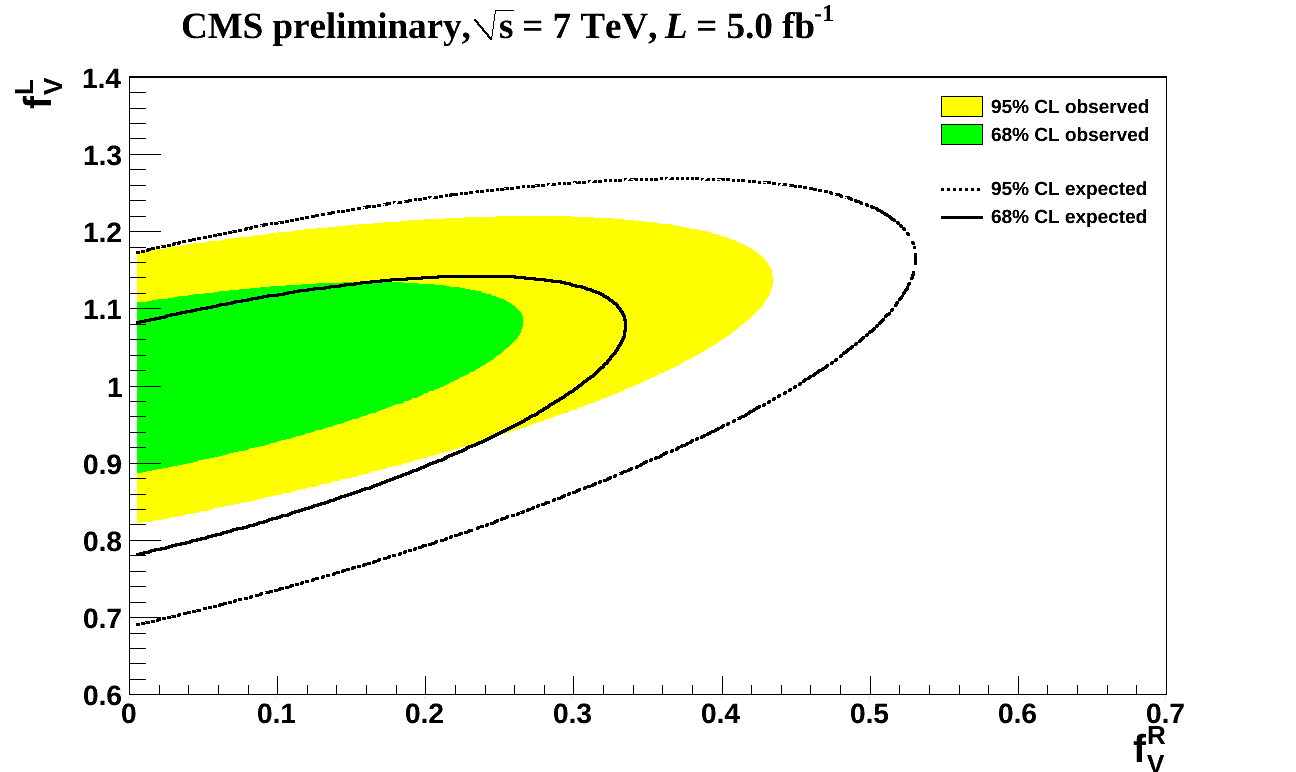}\quad
 \includegraphics[width=0.45\textwidth]{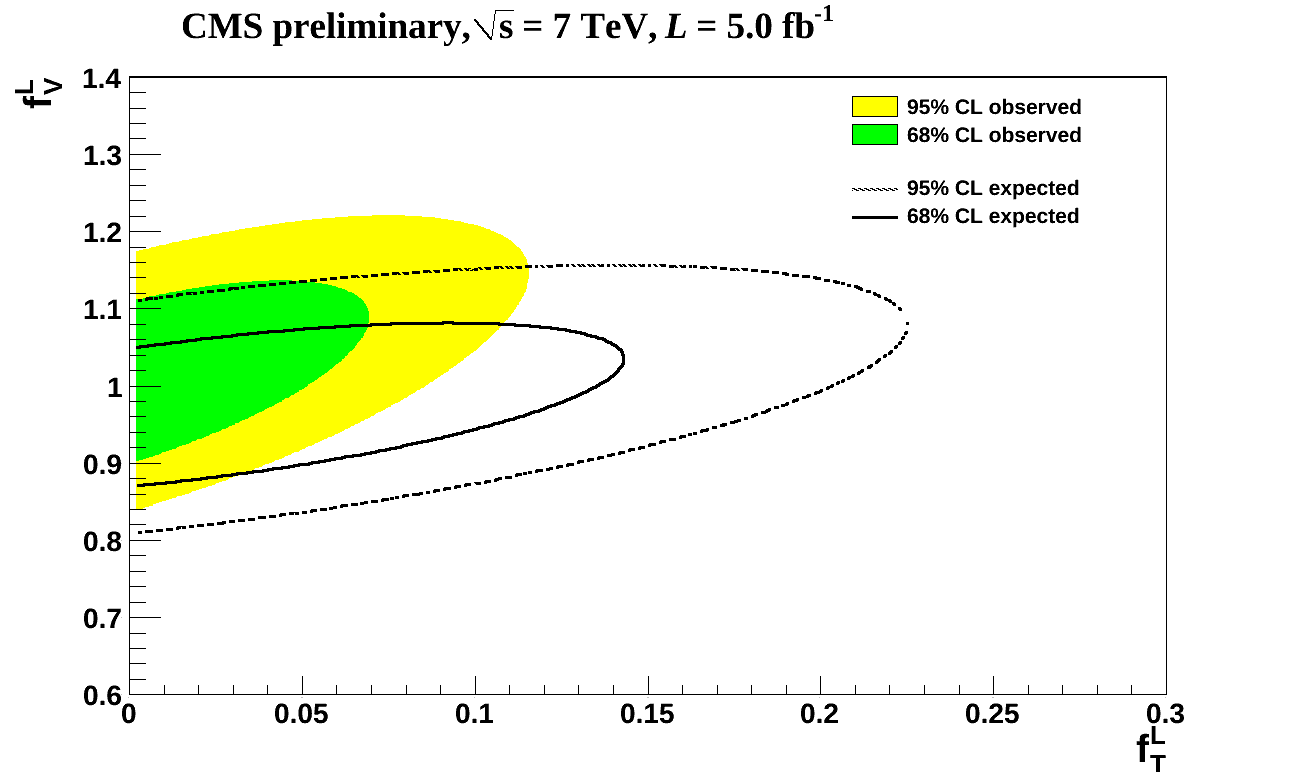}
\caption{Exclusion limits in the plane of left- and right-handed vector couplings (left) and left-handed vector and tensor couplings (right) in the $Wtb$ vertex.}
\label{FigWtbResults}
\end{figure}

A search for anomalous couplings in the $Wtb$ vertex was performed at $\sqrt s = 7$\,TeV explicitly accounting for anomalous contributions in both production and decay vertices~\cite{CMS:2014ffa}. Two scenarios were considered allowing for either right-handed vector~$f_V^R$ or left-handed tensor~$f_T^L$ anomalous couplings; in each case the left-handed vector coupling~$f_V^L$ was unconstrained and all remaining anomalous couplings were set to zero. An MVA-based analysis was conducted resulting in two-dimensional exclusion limits shown in Fig.~\ref{FigWtbResults}. They correspond to the following one-dimensional observed (expected) limits at 95\% confidence level: $|f_V^L| > 0.90$ (0.88) for $f_V^R = 0$, $|f_V^R| < 0.34$ (0.39) for $f_V^L = 1$, $f_V^L > 0.92$ (0.88) for $f_T^L = 0$, and $|f_T^L| < 0.09$ (0.16) for $f_V^L = 1$.

\section{Summary}

Measurements of single-top production cross sections in $t$ and $tW$~channels and a search for the $s$-channel production are presented. The value of $|V_{tb}|$ is extracted from the $t$-channel measurements. A number of properties of single-top production, namely $W$-boson helicity fractions, top-quark spin asymmetry, and anomalous couplings in the $Wtb$~vertex, are determined. All results are agreement with SM predictions.


\begin{thebibliography}{99}



\bibitem{Chatrchyan:2008aa}
  CMS Collaboration,
  \href{http://iopscience.iop.org/1748-0221/3/08/S08004}{JINST {\bf 3} (2008) S08004}.

\bibitem{LiuCKM14}
  Y.-F. Liu,
  to appear in the same proceedings.

\bibitem{Chatrchyan:2012ep}
  CMS Collaboration,
  JHEP {\bf 1212} (2012) 035
  [\href{http://arxiv.org/abs/1209.4533}{arXiv:1209.4533}].

\bibitem{Khachatryan:2014iya}
  CMS Collaboration,
  JHEP {\bf 1406} (2014) 090
  [\href{http://arxiv.org/abs/1403.7366}{arXiv:1403.7366}].

\bibitem{Chatrchyan:2012zca}
  CMS Collaboration,
  Phys.\ Rev.\ Lett.\  {\bf 110} (2013) 022003
  [\href{http://arxiv.org/abs/1209.3489}{arXiv:1209.3489}].

\bibitem{Chatrchyan:2014tua}
  CMS Collaboration,
  Phys.\ Rev.\ Lett.\  {\bf 112} (2014) 231802
  [\href{http://arxiv.org/abs/1401.2942}{arXiv:1401.2942}].

\bibitem{CMS:2013fmk}
  CMS Collaboration,
  \href{http://cds.cern.ch/record/1633190}{CMS-PAS-TOP-13-009}.

\bibitem{Khachatryan:2014vma}
  CMS Collaboration,
  \href{http://arxiv.org/abs/1410.1154}{arXiv:1410.1154} (submitted to JHEP).

\bibitem{CMS:2013rfa}
  CMS Collaboration,
  \href{http://cds.cern.ch/record/1601800}{CMS-PAS-TOP-13-001}.

\bibitem{CMS:2014ffa}
  CMS Collaboration,
  \href{http://cds.cern.ch/record/1702400}{CMS-PAS-TOP-14-007}.

\end{thebibliography}
\end{document}